\documentclass[letter]{aa}
\usepackage{natbib}
\usepackage{graphicx}
\usepackage{txfonts}
\bibpunct{(}{)}{;}{a}{}{,}

\newcommand{\msunyr}{\ensuremath{\mathit{M}_{\odot}{\rm yr}^{-1}}}   
\newcommand{\msun}{\ensuremath{\mathit{M}_{\odot}}}   
\newcommand{\mini}{\ensuremath{M_{\rm ini}}}                         
\newcommand{\lsun}{\ensuremath{\mathit{L}_{\odot}}}                  
\newcommand{\rsun}{\ensuremath{\mathit{R}_{\odot}}}                  

\newcommand{\lstar}{\ensuremath{\mathit{L}_{\star}}}                 
\newcommand{\mdot}{\ensuremath{\dot{M}}}                             
\newcommand{\mstar}{\ensuremath{\mathit{M}_{\star}}}                 
\newcommand{\rstar}{\ensuremath{\mathit{R}_{\star}}}                 
\newcommand{\teff}{\ensuremath{\mathit{T}_{\rm eff}}}                
\newcommand{\tstar}{\ensuremath{\mathit{T}_{\star}}}                 

\newcommand{\vrot}{\ensuremath{v_{\rm rot}}}                         
\newcommand{\vcrit}{\ensuremath{v_{\rm crit}}}                         

\newcommand{\tauross}{\ensuremath{\tau_{\mathrm{Ross}}}}                 

\newcommand{\mej}{\ensuremath{\mathit{M}_{\rm ej}}}   

\begin{document}

\title{Progenitors of supernova Ibc: a single Wolf-Rayet star as the possible progenitor of the SN Ib iPTF13bvn }

\author{Jose H. Groh\inst{1},  Cyril Georgy\inst{2}, and Sylvia Ekstr\"om\inst{1}}

\institute{
Geneva Observatory, Geneva University, Chemin des Maillettes 51, CH-1290 Sauverny, Switzerland; \email{jose.groh@unige.ch}
\and 
Astrophysics group, EPSAM, Keele University, Lennard-Jones Labs, Keele, ST5 5BG, UK
}

\authorrunning{Groh, Georgy, and Ekstr\"om}
\titlerunning{A single Wolf-Rayet star as the progenitor of iPTF13bvn}

\date{Received  / Accepted }

\abstract{Core-collapse supernova (SN) explosions mark the end of the tumultuous life of massive stars. Determining the nature of their progenitors is a crucial step towards understanding the properties of SNe. Until recently, no progenitor has been directly detected for SN of type Ibc, which are believed to come from massive stars that lose their hydrogen envelope through stellar winds and from binary systems where the companion has stripped the H envelope from the primary.
Here we analyze recently reported observations of iPTF13bvn, which could possibly be the first detection of a SN Ib progenitor based on pre-explosion images. Very interestingly, the recently published Geneva models of single stars can reproduce the observed photometry of the progenitor candidate and its mass-loss rate, confirming a recently proposed scenario. We find that a single WR star with initial mass in the range 31--35~\msun\ fits the observed photometry of the progenitor of iPTF13bvn. The progenitor likely has a luminosity of $\log (\lstar/\lsun)\sim5.55$, surface temperature $\sim45000~K$, and mass of $\sim10.9~\msun$ at the time of explosion. Our non-rotating 32~\msun\ model overestimates the derived radius of the progenitor, although this could likely be reconciled with a fine-tuned model of a more massive (between 40 and 50~\msun), hotter, and luminous progenitor. Our models indicate a very uncertain ejecta mass of $\sim8~\msun$, which is higher than the average of the SN Ib ejecta mass that is derived from the lightcurve (2--4~\msun). This possibly high ejecta mass could produce detectable effects in the iPTF13bvn lightcurve and spectrum. If the candidate is indeed confirmed to be the progenitor, our results suggest that stars with relatively high initial masses ($>30~\msun$) can produce visible SN explosions at their deaths and do not collapse directly to a black hole.}

\maketitle

\section{\label{intro}Introduction} 

\defcitealias{gmg13}{G13}
\defcitealias{eldridge13}{E13}
\defcitealias{cao13}{C13}

Core-collapse supernova (CCSN) explosions mark the end of the life of stars with initial masses (\mini) above $\sim8~\msun$. CCSNe are generally luminous events that are classified according to their spectrum and lightcurve (\citealt{fillipenko97}), with type II SNe presenting strong H lines in their spectrum. Conversely, type I CCSNe do not present H lines and are further subdivided into types Ic (without He lines) and Ib (with He lines).

Most of the observed features of SNe, such as the temporal evolution of the spectrum, absolute magnitudes, and colors, are determined by characteristics that can be traced back  to properties of the progenitor, such as its mass, radius, and chemical composition. For that reason, it is thought that different kinds of SNe are produced by different progenitors, which in general would come from both single and binary stars.

The properties of SN progenitors can be constrained using several complementary techniques. These include direct imaging of the progenitor in pre-explosion images \citep[e.g.][]{smartt09a}, hydrodynamical modeling of the lightcurve  \citep[e.g.][]{nomoto93} and spectrum \citep[e.g.][]{dessart11b}, and analysis of the progenitor's environment \citep[e.g.][]{modjaz08}, among others. To extract all the information about the progenitor contained in the observations, these techniques are often used in combination with stellar evolution models.

While the progenitors of SN II have fortuitously been directly imaged in the past decades \citep{smartt09a}, no detection of a SN Ibc progenitor has been obtained until recently (\citealt{smartt09a}; \citealt{eldridge13}, hereafter E13). Supernova Ibc progenitors are believed to come in part from massive stars that lose their H envelope due to stellar winds, becoming Wolf-Rayet (WR) stars \citep{maeder81}, and in part from binary systems where the H envelope has been stripped by a companion \citep{pod92}. However, the non-detection of WRs as SN Ibc progenitors has cast doubt on whether single WR stars could produce SN Ibc. While \citet{smartt09a} and \citetalias{eldridge13} suggest that single WRs are bright and should have been detected in pre-explosion images, \citet{yoon12} suggest that single WR stars at the pre-SN stage could be very hot and faint in the optical bands.  Based on detailed evolutionary and atmospheric modeling, \citet{gmg13} (hereafter G13) confirm that SN Ibc progenitors from single stars are too faint to have been detected in the pre-explosion optical images of the SN Ibc sample from \citetalias{eldridge13}. Therefore, not detecting WR stars in SN pre-explosion images do not preclude them from being the progenitors of SN Ibc.

Recently, \citet{cao13} (hereafter C13) report a possible detection of the progenitor of the SN Ib iPTF13bvn. The candidate source was detected within the 2$\sigma$ error of the SN astrometry and, based on its absolute magnitudes, has been suggested as a massive WR star, with a hydrostatic radius smaller than a few \rsun\ and a mass-loss rate (\mdot) of $\sim3\times10^{-5}\msunyr$ \citepalias{cao13}.

\begin{figure*}
\center
\resizebox{0.995\hsize}{!}{\includegraphics{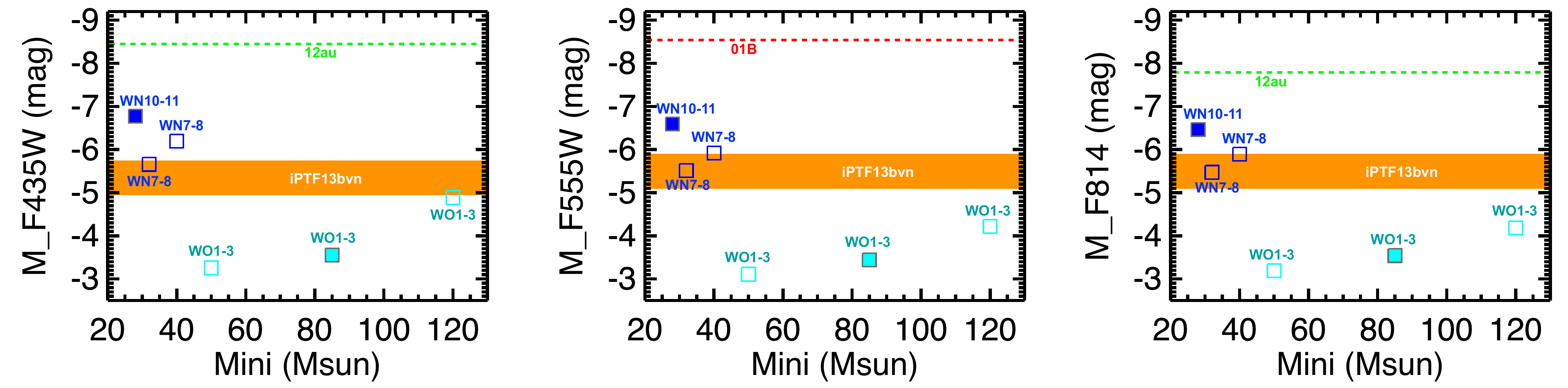}}
\caption{\label{snibprog} {Absolute magnitude of the SN Ib progenitor models as a function of $\mini$, in the ACS/WFC  $F435W$ (left), $F606W$ (middle), and $F814W$ filters (right). The progenitor spectral type is color coded, with cyan symbols corresponding to WOs and blue symbols to late WN stars. The evolutionary model type is coded as open (without rotation) and filled (with rotation) symbols. The orange regions correspond to the absolute magnitudes of iPTF13bvn derived from the measured photometry. The upper limits of other SN Ib progenitors (2001B and 2012au, from \citetalias{eldridge13}) are shown by horizontal dashed lines. The non-rotating 32~\msun\ model fits the observations of the iPTF13bvn progenitor relatively well.}}
\end{figure*}

Our goal here is to investigate the nature of the possible progenitor of iPTF13bvn, and constrain the luminosity, surface temperature, spectral type, and initial mass of the progenitor. This is achieved by comparing theoretical predictions of SN progenitors, based on combined stellar evolution and atmospheric models from \citetalias{gmg13} to the observed properties of iPTF13bvn from \citetalias{cao13}.

\section{\label{model}Stellar evolution and atmospheric models of SN progenitors}

Our strategy here is to employ our recent Geneva stellar evolution and atmospheric models of stars \mini= 9--120~\msun, computed in \citet{gme13} and \citetalias{gmg13}. Therefore, the models presented here are not fine-tuned to reproduce any specific observation. Below we briefly remind the reader of the main properties of the models and refer to the aforementioned papers for further details.

The evolutionary models were computed with the Geneva stellar evolution code and are presented in their vast majority in \citet{ekstrom12}. A few additional models are shown in \citetalias{gmg13}. The models assume solar metallicity\footnote{The host galaxy of iPTF13bvn (NGC5806) has an oxygen abundance of $12+\log(N_\mathrm{O}/N_\mathrm{H})=8.5$ \citep{smartt09b}, so the use of models at solar metallicity is appropriate.}  and were computed both for the case of no initial rotation and for an initial rotational speed of 40\% of the critical velocity on the zero-age main sequence.
The output spectra are computed using the atmospheric radiative transfer code CMFGEN \citep{hm98}, which is a state-of-the-art, full line-blanketed, non-local thermodynamical equilibrium, spherically symmetric code. As input parameters, the models computed here assume the physical conditions provided by the stellar evolutionary calculations, such as the hydrostatic radius \rstar, luminosity \lstar, mass \mstar, and surface abundances. The stellar evolution solution and the atmospheric solution are merged using the temperature structure of the envelope. A hydrostatic solution is computed for the subsonic atmosphere and is joined to a $\beta$-type wind velocity law. Our models consider wind clumping with a volume-filling factor of $f=0.1$.

Our grid of models allows one to employ, for the first time, theoretical synthetic spectra and photometry of SN progenitors that were computed self-consistently with the evolutionary calculations to analyze the observations of SN progenitors. 

\section{\label{snib} The elusive progenitors of SN Ibc and the case of iPTF13bvn}

The nature of the SN Ib progenitors from single stars depends on their \mini\ and rotation \citepalias{gmg13}. We do not expect that our models are able to reproduce all SN Ib progenitor observations, but only those arising from single stars. We briefly recall the main findings from \citetalias{gmg13} concerning the SN Ib progenitors from single stars (see their Fig. 4).  Non-rotating models predict that 72\% of the progenitors are late WR stars of the WN subtype (WN7--8) and have $32.0~\msun < \mini < 45.0~\msun$. The remaining 28\% of the SN Ib progenitors are WR stars of the WO subtype (WO1--3) with $45.0~\msun < \mini < 52.2~\msun$ or  $106.4~\msun < \mini < 120.0~\msun$. In this context, the non-rotating models are representative of stars that are born as slow rotators. Rotating models with $\vrot/\vcrit=0.4$ indicate that 96\% of the SN Ib progenitors are WN 10--11 stars ($25.0~\msun < \mini < 30.1~\msun$), while 4\% are WO 1--3 stars ($82.0~\msun < \mini < 88.7~\msun$). These mass ranges assume chemical abundance criteria to determine the SN type, with SN Ib having no H and more than 0.6~\msun\ of He in the ejecta, and SN Ic having no H and less than 0.6~\msun\ of He. As discussed in, say, \citet{georgy12a},  the mass ranges quoted above are uncertain.

Our models produce several quantities that can be compared to data derived from the observations of SN progenitors, allowing one to constrain the nature of the progenitor. 
We argue that absolute magnitudes are the most direct property that can be employed in the comparison, since they are obtained from the observations of SN progenitors in an almost model-independent way. 

Figure~\ref{snibprog} presents the absolute magnitudes of our SN Ib progenitors as a function of \mini. These broadband absolute magnitudes are computed by convolving the high-resolution spectrum of the progenitor with a given filter bandpass.  As such, they include both the contributions of spectral lines and continuum, since both are probed by the observed broadband photometry. In the present case, we use the same filters in which the possible iPTF13bvn progenitor was detected ({\it Hubble Space Telescope}/Advanced Camera for Surveys, {\it HST}/ACS). This illustrates one of the strengths of our approach, since no filter transformations between different photometric systems are needed.

\begin{figure*}
\center
\resizebox{0.78\hsize}{!}{\includegraphics{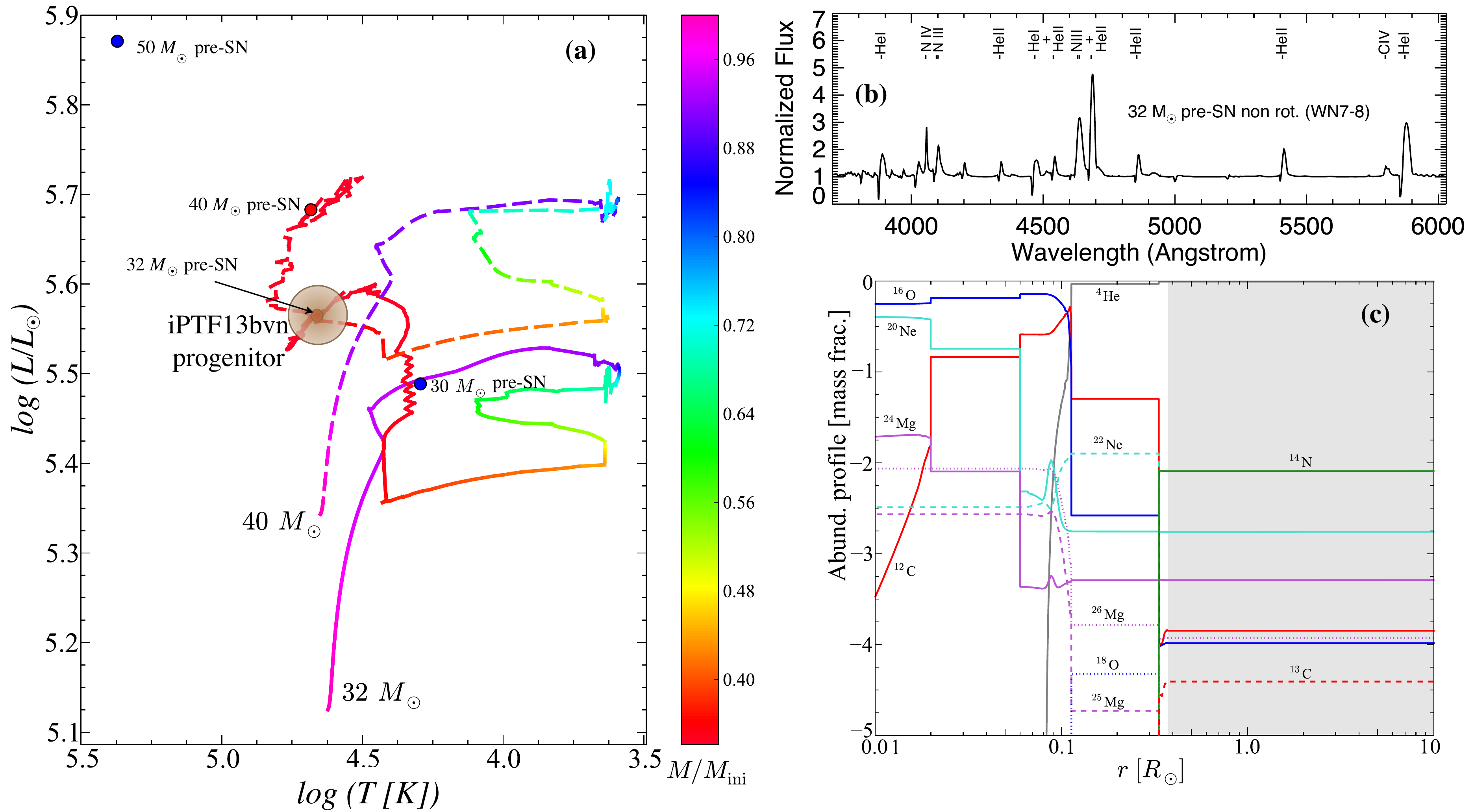}}
\caption{\label{abund} {{\it (a)}: HR diagram showing the evolutionary tracks of non-rotating 32~\msun\ (solid) and 40~\msun\  models (dashed), with the red circles representing their endpoints. The location of the possible progenitor of iPTF13bvn is depicted by the brown region, whose size does not correspond to a formal error analysis and is merely indicative of the possible location of the progenitor. The color code refers to the mass at each timestep divided by \mini. The endpoints of the non-rotating 30 and 50~\msun\ models are also shown (blue circles).  {\it (b)}: Continuum-normalized optical spectrum of the 32~\msun\ model at the pre-SN stage, showing it is a typical WN7--8 star (\citetalias{gmg13}). The strongest features are identified. {\it (c)}: Chemical structure of  our 32~\msun\ model at the end of the core carbon-burning stage, when the model has 10.9~\msun. The variations in the mass fraction of different species is shown as a function of the radius, with the surface of the star being to the right, at 9.69~\rsun, and the center to the left at 0.0~\rsun.}}
\end{figure*}

Our progenitor models can be divided into two main groups: relatively bright ($M_{\it 555W}$=$-5.5$ to $-6.2$ mag) massive WNLs,  and relatively faint ($M_{\it 555W}$=$-3.1$ to $-4.1$ mag), very massive WOs. For other well-classified SN Ib, such as SN 2001B and SN 2012au, all models lie below the detection threshold, in agreement with a non-detection in the pre-explosion images \citepalias{gmg13}. To avoid errors due to filter transformations, we decided to use the original {\it HST}/ACS photometry of iPTF13bvn from \citetalias{cao13} to obtain the absolute magnitudes in these filters, instead of using the absolute magnitudes in $BVI$ filters provided by these authors. They measured $m_{\it 435W}=26.7 \pm 0.2$~mag, $m_{\it 555W}=26.5 \pm 0.2$~mag, and $m_{\it 814W}=26.4 \pm 0.2$~mag. We also assume the same distance (d=22.5~Mpc; \citealt{tully09}), and reddening (local extinction $E(B - V )$ = 0.0437, foreground extinction $E(B - V )$=0.0278, and $R_V$=3.1) estimated by \citetalias{cao13}. We thus find that the possible progenitor has $M_{\it 435W}=-5.3 \pm 0.4$~mag,  $M_{\it 555W}=-5.5 \pm 0.4$~mag, and $M_{\it 814W}=-5.5 \pm 0.4$~mag.

We find that some of the non-rotating WNL SN Ib progenitor models are in excellent agreement with the observed photometry  (Fig. \ref{snibprog}). This is remarkable given that the models have not been fine tuned. For instance, the non-rotating model with \mini=32~\msun\ has $M_{\it 435W}$=$-5.66$~mag,  $M_{\it 555W}$=$-5.52$~mag, and $M_{\it 814W}$=$-5.47$~mag.  The non-rotating 40~\msun\ model is brighter than the observations in the $F435W$. The 50~\msun\ model is too faint in all bands in which the progenitor was detected. SN Ib progenitors with $\mini>50~\msun$  are WO stars (cyan) and are typically at least 1~mag fainter than the observations.  SN Ib progenitors with $\mini=28~\msun$ are too bright and can also be discarded. SN progenitors with $\mini< 25~\msun$ do not produce SN Ib \citep{georgy12a}. Thus, while our non-rotating 32~\msun\ model fits the observed photometry of the possible progenitor of iPTF13bvn best, we cannot discard that a fine-tuned non-rotating model between 40--50~\msun\ (or a rotating model between 28 and 32~\msun) could also fit the data (Fig. \ref{snibprog}). We note that the absolute magnitudes of SN Ib progenitors are strongly dependent on \mini\ (Fig. \ref{snibprog}), which is the reason \mini\ is well constrained. 

Therefore, if the detected star is indeed the progenitor of the SN Ib iPTF13bvn, our models indicate that the progenitor is consistent with a single WR star that had $\mini\sim$31--35~\msun. Figure \ref{abund}a shows the evolutionary tracks of the non-rotating 32 and 40~\msun\ models in the Hertzprung-Russell (HR) diagram. The progenitor of iPTF13bvn would have a final mass of 10.9~\msun, surface temperature of $\tstar\sim45000~$K (computed at a Rosseland optical depth of $\tauross=20$), effective temperature of $\sim40000~$K (computed at $\tauross=2/3$), and $\log (\lstar/\lsun)\sim5.55$. According to our models, the progenitor would have a spectral type of WN 7--8o, with the optical spectrum characterized by broad, strong emission lines of \ion{He}{ii},  \ion{He}{i},  \ion{N}{iii}, and  \ion{N}{iv} (Fig. \ref{abund}b). 

Our models also predict other theoretical quantities that can be compared to those derived from the observations. We stress here that unlike the progenitor photometry, these other quantities are obtained from the observations by modeling the SN lightcurve and/or spectrum. In the case of iPTF13bvn, there are two other physical properties that have been determined (see \citetalias{cao13}), namely the progenitor hydrostatic radius (less than a few \rsun\ ) and $\mdot\sim3\times10^{-5}\msunyr$. Regarding \mdot, our non-rotating 32~\msun\ model predicts that the SN progenitor will have $\mdot=2.9\times10^{-5}\msunyr$, in excellent agreement with the value derived from the observations. The hydrostatic radius of  this model, however, is 9.65~\rsun, which seems to be larger than what is allowed from the lightcurve modeling ($<$ few \rsun; \citetalias{cao13}). A fine-tuned model with \mini\ between 40 and 50~\msun\ could probably yield a lower value for the hydrostatic radius and still fit the observed photometry and mass-loss rate.

The chemical abundances and total amount of mass in the SN ejecta (\mej)  also provide constraints on the nature and chemical structure of the progenitor star itself.  Figure \ref{abund}c presents the chemical structure of the 32~\msun\ at the pre-SN stage, which we suggest is similar to the progenitor of iPTF13bvn. This model has no H left because of mass loss due to stellar winds, and the CO core has 8.9~\msun. As of yet, no determination of $\mej$ of iPTF13bvn exists in the literature.  Our 32~\msun\ model predicts $\mej=7.89~\msun$, of which 2.89~\msun\  are composed of He \citep{georgy12a}. We note here that while our models are able to provide a good estimate for the amount of He in the ejecta, $\mej$ may be quite uncertain due to difficulties in estimating the remnant and fallback mass. Based on 1-D hydrodynamical models, \citet{ugliano12} suggest a remnant mass of $\sim2.0~\msun$ for 32~\msun\ stars, implying $\mej\sim9.0~\msun$. Although uncertain, this value of $\mej$ is much higher than the average $\mej$ of SN Ib that is usually derived from the SN lightcurve ($\sim2-4\msun$; \citealt{drout11, cano13}). This would imply that the high $\mej$ could have detectable effects on the SN lightcurve and spectrum, making iPTF13bvn a peculiar SN Ib. Because the high luminosity of the candidate progenitor implies a high mass, a large $\mej$ would be inferred independently of whether the progenitor followed a single or a binary evolution scenario.

\section{\label{noprog} What if the candidate star is not the progenitor of iPTF13bvn?}

As discussed by \citetalias{cao13}, imaging of the SN site in a few years time is needed to confirm that the candidate star is indeed the progenitor of iPTF13bvn.  We now assume that the source detected close to the SN position is not the progenitor of iPTF13bvn and discuss the possible scenarios.

First, a more massive WR of the WO spectral type, with $\mini=50$ to 120~\msun, could be undetectable in the pre-explosion images (Fig. \ref{snibprog}). Assuming that the putative undetected progenitor has similar amounts of reddening as the one towards the candidate, these WOs would appear with $m_{\it 435W}=27.5$ to 28.9 mag, $m_{\it 555W}=27.8$ to 28.9 mag, and $m_{\it 814W}=27.7$ to 28.7 mag at the distance of iPTF13bvn. They would most certainly be fainter than the magnitude limit of the pre-explosion observations, unless the latter are unusually deep. Since the specific values  of the magnitude limits in the different bands are not available to us, we cannot gauge the likelihood of this scenario. 

Second, we cannot leave out the possibility that a reddened WN star similar to the one discussed above, with $\mini$=32~\msun, is the actual progenitor. A reddening $\sim$1--2 mag  greater than the local reddening of $E(B-V)=0.0437$ estimated by \citetalias{cao13}, perhaps because of the presence of circumstellar material, would make the WN star undetectable. In this case, the progenitor would have the same nature as we suggested in Sect. \ref{snib}.

Third, a low-mass, faint WR star formed via binary evolution could be the progenitor, as expected from binary models \citep[e.g.,][]{pod92}. These progenitors would in principle be much fainter (because they have low mass) than the detection limit of the pre-explosion images. For instance, low-mass binary stars known as V Sagittae stars \citep{steiner05}, such as HD 45166 \citep{groh08},  have been suggested as SN Ibc progenitors \citepalias{eldridge13}. As shown by \citet{groh08}, these stars have relatively low luminosities ($\log (\lstar/\lsun)=3.75$) and high effective temperatures (\teff=50000 K), so they would be undetectable in the pre-explosion images. However, it would be challenging to reconcile the $\mdot$ values found for the progenitor of iPTF13bvn ($\sim3\times10^{-5}\msunyr$) with the values found for low-mass, WR-like stars that had their envelope stripped by a companion such as V Sag stars. They have $\mdot\sim2\times10^{-7}\msunyr$ \citep{groh08}, which is about  two orders of magnitude lower than is inferred for the progenitor of iPTF13bvn. In addition, the companion of these stars are cool, late B-type stars with a comparably low $\mdot$, so the companion could not be responsible for the relatively high $\mdot$ inferred for the progenitor.

\section{\label{prospects} Prospects for detecting SN Ibc progenitors}

We have shown that the photometry of the candidate progenitor of the SN Ib iPTF13bvn is consistent with that of a single WR star at the pre-SN stage, confirming the results from \citetalias{cao13}. Here, we went one step further and estimated that the progenitor has a WN~7--8 spectral type and  $\mini\sim31-35~\msun$. Since SN Ib comprises 35\% of the total sample of SN Ibc, one may wonder to what extent it is likely that a SN Ib progenitor has possibly been detected, while many SN Ic progenitors remain undetected in pre-explosion images. Our models indicate that most of the SN Ib progenitors  (70--96\%) are brighter than the SN Ic progenitors \citepalias{gmg13}. This is because the majority of SN Ib progenitors are not as hot as the SN Ic progenitors, so the earlier ones have more flux in the optical bands. Because of that, while SN Ibs are rarer than SN Ics, it is more likely that a SN Ib progenitor is detected in pre-explosion images, and SN Ib progenitors will likely dominate future samples of directly detected progenitors of SN Ibc.

We can also estimate the maximum distance up to which a SN Ib progenitor coming from single-star evolution would be directly detected in pre-explosion images. This depends on the absolute magnitude and reddening of the progenitor in the filter of the pre-explosion image and on the magnitude limit of the image. For this purpose, we use the absolute magnitudes from Fig. \ref{snibprog} and assume a magnitude limit of 26.7 mag in F435W and low reddening (similar to the one towards the progenitor of iPTF13bvn).  Since the brighter SN Ib progenitor according to out models is the one from the rotating 28~\msun\ model, we find that SN Ib progenitors could be detected up to 42~Mpc.

It is interesting to compare this maximum distance for detecting SN Ib progenitor with the one derived for single SN Ic progenitor models. For the same F435W filter, magnitude limit, and reddening, a value of 8.4~Mpc is obtained \citepalias{gmg13}. Thus, SN Ib progenitors should be detected up to much greater distances than SN Ic progenitors. In addition, for distances over a certain threshold (8.4~Mpc given the assumptions above), the sample of directly detected SN progenitors will be incomplete, with the majority of the sources being SN II and a few being SN Ib.

The possible detection of a SN Ib progenitor is also extremely interesting since it indicates that a very massive progenitor ($\mini > 30~\msun$) can still produce a visible SN event. If a black hole (BH) has been produced, then this would also indicate that a BH formation can sometimes be linked to a visible SN event. Of course, this last conclusion is quite speculative, since it is fairly difficult to know if the iPTF13bvn remnant is a BH or NS. However, we cannot discard that this question may be answered in the future, and efforts towards that goal are warranted.

\begin{acknowledgements}
We thank Georges Meynet for many interesting discussions and comments, and John Hillier for making CMFGEN available. JHG is supported by an Ambizione Fellowship of the Swiss National Science Foundation. CG acknowledges support from EU- FP7-ERC-2012-St Grant 306901. 

\end{acknowledgements}
\bibliographystyle{aa}
\bibliography{../../refs}

\end{document}